\documentclass[english]{elsart} 
\usepackage[T1]{fontenc} 
\usepackage[latin1]{inputenc} 
\usepackage{amsmath} 
\usepackage{graphicx} 
\usepackage{setspace} 
\usepackage{epsfig} 
\makeatletter 
 
\providecommand{\LyX}{L\kern-.1667em\lower.25em\hbox{Y}\kern-.125emX\@} 
 
\usepackage{babel} 
\makeatother 
\begin{document} 
\begin{frontmatter} 
\title{Invisible Hand Effect in an Evolutionary Minority Game Model}

\author{Marko Sysi-Aho,} 
\author{Jari Saramäki,} 
\author{Kimmo Kaski} 
 
\address{Laboratory of Computational Engineering, Helsinki 
University of Technology, \\ 
P. O. Box 9203, FIN-02015 HUT, Finland.} 
 
\begin{abstract} 
In this paper we study the properties of a Minority Game 
with evolution realized by using genetic crossover to modify 
fixed-length decision-making strategies of agents. 
Although the agents in this evolutionary game act selfishly by  
trying to maximize their own performances only, it turns out  
that the whole society will eventually be rewarded optimally.  
This "Invisible Hand" effect is what Adam Smith  over two  
centuries ago expected to take place in the context of free  
market mechanism. However, this behaviour of the society of agents  
is realized only under idealized conditions, where all agents are 
utilizing the same efficient evolutionary mechanism. If on the  
other hand part of the agents are adaptive, but not evolutionary,  
the system does not reach optimum performance, which is also  
the case if part of the evolutionary agents form a uniformly  
acting "cartel". 
\end{abstract} 
\end{frontmatter} 
\maketitle 
 
\section{Introduction} 
 
In his book of 1776 Adam Smith outlined a mechanism which he  
supposed to describe the behaviour of economic societies \cite{smith}.  
He postulated that individuals who try to maximize their own gain   
without active regard to the society's welfare will eventually  
reward the society most effectively. As the mechanism how this 
should actually happen Smith described it as  an ``invisible hand''  
of a benevolent deity administering human happiness by leading  
individuals to act in a certain way. 
In the modern context invisible hand processes have been studied  
as part of Game Theory, a branch of mathematics dealing with payoffs 
and strategies, where the interrelationships between the best  
productivity of individual actors and the society has been refined  
by John Nash through equilibrium concept \cite{gibbons,fudenberg,rustichini}. 
He indicates that individuals could only maximize their own benefit 
by taking other individuals into account. 
However, Smith's assumption about the optimal performance of the society  
through selfish individuals turns out to be valid in certain circumstances. 
For example this is the situation for the Minority Game introduced by Challet  
and Zhang ~\cite{challet1}, see also 
Refs.~\cite{internet,zhang,savit,manuca,johnson,cavagna,challet2}. 

Minority games are repeated coordination games \cite{gibbons,fudenberg} 
where agents use a number of different strategies in order to join one of 
the two available groups, A or B, and those who belong to the minority 
group are rewarded. In the original MG \cite{challet1} the agents are 
exposed to $P$ different histories and the strategy of an agent determines 
the choice of the group for each history. Thus, the length or dimension 
of a strategy equals $P$, and the set of all possible $2^P$ strategies 
composes a strategy space from which the agents' strategies are randomly 
drawn in the beginning of the game. Strategies are cumulatively scored 
based on correct minority group choices, and at each step of the game the  
choices of the agents are determined by their highest-scoring  
strategies. In the following, we shall refer to this basic Minority 
Game with the above described adaptation mechanism as BMG, and use  
the abbreviation MG to refer to the Minority Game concept in a  
more general fashion.  
 
Minority games can be viewed as simulating the performances of 
competing individuals and the welfare of the society they compose. This  
kind of mechanism could coarsely speaking be involved in a stock market 
where investors share information and make buy-or-sell decisions 
in order to gain profit. If the number of sellers of a particular stock  
is larger than the number of buyers, supply exceeds demand and one  
expects a decrease in the stock price \cite{parkin}. Then the buyers,
being in minority,  would win due to the low price levels. 
In the opposite case sellers would win, because excess demand  
would increase the price of the stock. In the long run, the price of the  
stock eventually settles down to its equilibrium value, i.e. supply  
and demand are, on average, close to each other and the public  
information has been efficiently utilized. In relation to this the  
utility or performance of the society can be viewed as the number of  
content individuals. In other words if everybody agrees on the price,  
both the sellers and buyers are content. In the framework of MG,  
this means that the numbers of buyers and sellers are as close to  
each other as possible and the game is in one of its pure-strategy  
Nash equilibria \cite{gibbons,fudenberg,bot,cmz,marsili}.
On the other hand if the numbers deviate from equilibrium, either one 
of the groups is dissatisfied and thus the overall "happiness" of 
the society decreases.  
 
In these games the long-term system performance is of major interest. 
It is measured as the variation of the minority group size around its  
maximum, such that the larger the minority group size at every time step 
is, the better the aggregate system performance is. Usually, the behavior  
of the system performance depends on a control parameter $z$ \cite{savit},  
which combines the dimension of the strategy space and the number of  
players in the game. In the BMG the best system performance occurs 
at $z=z_c$, which depends on the number of the strategies each agent 
has \cite{challet4}. On the other hand, applying an evolutionary mechanism 
to the agents' strategies usually changes the behaviour of the game 
remarkably. Coarsely speaking, the evolutionary mechanisms studied 
in the context of minority games can be divided into two groups, i.e. to 
those mechanisms that are applied to pure strategies and do not explicitly 
include probabilities in strategy selection or decision-making of the agents 
(e.g. Ref.~\cite{savitI}), and those that do so  
(e.g. Refs.~\cite{cmz,marsili,reents,johnsonII}).  
A further division can be made between fixed (e.g. Ref.~\cite{savitI}) 
and variable-length strategies (e.g. Ref.~\cite{savitII}). 
 
Our evolutionary minority game belongs to the pure-strategy 
class with fixed-length strategies. However, the main difference  
between our game and the game discussed in Ref.~\cite{savitI}, 
belonging also to the same class, is the genetic-algorithm-based 
mechanism by which the strategies of an agent are modified. 
We find that enhancing the BMG with one-point genetic crossover mechanism 
results in the birth of new strategies based on well-performing 
parent strategies and leads to behavior resembling Smith's ``invisible hand''. 
Previously we have studied the effect of genetic crossover of  
strategies on the MG performance ~\cite{marko1,marko2,marko3}, and  
shown that our simple pure-strategy evolutionary mechanism 
leads to highly enhanced performance, both at the system as well  
as at the individual agent level. Recently Yang et al.~\cite{yang} 
have reported results of a study using a genetic-algorithm-based 
evolutionary mechanism, which turned out to be quite similar to 
those of ours~\cite{marko1,marko2,marko3}. 
Below we will show that with our evolutionary mechanism the optimal  
system performance can be reached for a wide range of control  
parameter values. In contrast to Ref.~\cite{savitI}, increasing  
the number of strategies increases the system performance.  
Furthermore, the optimal performance is typically reached for all  
possible histories independent of their order of appearance, as  
the histories are randomly drawn from a uniform distribution in  
order to avoid any repetitive history cycles.  
 
This paper is organised such that first we introduce our evolutionary 
minority game (EMG) model. Then we show simulation results on the 
system performance and compare them with the optimal limit as well 
as with results of simulations using the BMG. Furthermore, we 
investigate using the Minimum and Maximum Spanning Tree methods 
whether similarly 
performing evolutionary agents form clusters in the sense that 
they would play similar strategies, and whether well performing
agents' strategies tend to be different from those of the badly performing ones.
In addition, we briefly discuss 
the effect of a fraction of the agents forming a uniformly acting 
group, "cartel", on society utility. Finally we draw conclusions.  
  
\section{Model} 
 
Let us first briefly describe the BMG, and then discuss the strategy 
evolution method we have applied. The BMG \cite{challet1} consists 
of (odd) N  agents who simultaneously choose between two options, 
denoted $1$ and $-1$.  After the decisions of the agents, votes are 
counted, and those who belong to the minority group gain profit. 
The winning minority is publicly announced after every round. The 
game is repeated, and at each round the choice of an agent is determined 
by a component of a $P$-dimensional binary vector called the strategy 
of the agent. Each of the $P$ components indicates a response 
corresponding to a particular history vector of length $M$,   
which comprises of the minority choices during the last $M$ rounds. As 
there are $2^M$ possible histories, $P=2^M$ \cite{challet1}.  
 
In the BMG histories are explicitly determined by the choices 
of the agents, but instead we have decided to draw the histories 
randomly from a uniform distribution. The motivation for this was  
to avoid occurrences of any cyclic patterns of repeating histories,  
thus pre-empting the agents' possibilities for history-pattern-based 
coordination.  Furthermore, previous results \cite{marko1} with  
deterministic histories show that with our evolutionary 
mechanism, the game would finally repeat a single history only.  
The method of randomly selecting histories provides a 
stronger basis to justify the success of the chosen evolutionary 
mechanism in explaining the observed highly efficient system 
performance. (For discussion on the effect of using random versus 
non-random histories, see e.g.~\cite{cavagnaI,challet,lee}). 
In the BMG each agent has $S$ randomly chosen strategy vectors  
$s_i$ that are scored according to their cumulative success in  
predicting the minority group, with unit score added 
for the right choice and deducted in the opposite case.  
At each round an agent uses the strategy vector $s_i$ with the  
highest score. 
 
We define the performance of an agent at each round to be the number 
of times it has belonged to the minority minus the number of times it 
has belonged to the majority, and then scaled into the interval $[0,1]$. 
In order to measure characteristics of the whole system of agents 
we define the society utility $u(t)\in [0,1]$ at each round $t$ to 
be the number of agents who belong to the minority group divided by 
$(N-1)/2$ (the maximum size of the minority group). In the minority 
game studies a common measure to characterise the model is the 
attendance \cite{challet4}  
 
\begin{equation} 
a(t) = \sum_{i=1}^N \sigma_i(t),  
\label{attendance} 
\end{equation} 
 
where $\sigma_i(t) \in \{ -1, 1 \}$ denotes the action which the 
agent $i$ takes at round $t$. Thus the attendance gets values  
$a \in \{ -N,-N+2,...,1,-1,...,N-2,N \}$ and it is related to the  
society utility as 
 
\begin{equation} 
u(t) = \frac{N-|a(t)|}{N-1}. 
\label{society_utility} 
\end{equation} 
 
If $a=1$ or $a=-1$, the society utility is at its maximum $u=1$. When  
$a$ increases, the society utility $u$ decreases. In the minority game 
studies it is a common practice to observe the normalized fluctuations 
of attendance  
 
\begin{equation} 
\left<a^2\right>/N = \frac{1}{NT}\sum_{t=k+1}^{k+T}a²(t) 
\label{fluctuation} 
\end{equation} 
 
as function of the control parameter $z=2^M/N$, see \cite{savit}. 
The square of attendance Eq.~(\ref{attendance}) is averaged 
over $T$ time-steps and then normalized by the number of agents. 
In the BMG with a fixed number of strategies $S$ per agent one 
can separate three regions in the normalized fluctuations as $z$ 
changes: for small values of $z$ fluctuations are large, 
for intermediate values of $z$  
they reach a minimum, and for large values of $z$ they start to  
converge towards the limit of random decisions, being unity  
(decisions taken by flipping a coin) \cite{challet4}.  
According to Eq.~(\ref{society_utility}), small normalized fluctuation 
values indicate large values of the society utility. If we increase  
$S$ in the BMG, normalized fluctuations also increase and thus  
the society utility decreases. As we will see later, the behaviour  
is very different in our EMG: increasing $S$ leads to larger  
society utility values, and separate regions of fluctuation levels do not  
exist. 
 
In contrast to the BMG \cite{challet1}, where the strategies remain the same  
throughout the game, we utilize an evolutionary mechanism that allows  
agents to change their strategies for better personal gain.  
This mechanism is as follows: after every $r$ rounds the agents observe  
the performances of their neighbours, and if they are doing worse  
than a neighbour, they cross two of their best $S$ strategies 
and replace the two worst strategies with the resulting ones.  
The crossover  
is done in a typical genetic algorithm fashion 
\cite{goldberg,lawrence}: a crossover point $p_c \in [0,P]$, is 
randomly selected, and the children  
inherit $p_c$ strategy components from one parent and $P-p_c$ from the other. 
For example, if the parent vectors were $(1\  1\  1\  1 )$ 
and $(-1 -1 -1 -1)$ and the crossover point $p_c=2$, the 
resulting vectors would be $(1\  1 -1 -1)$ and $(-1 -1\  1\  1)$.  
 
The rule of when and which agents will attempt to improve their  
strategies can be implemented in many ways. The only aim of  
determining the rule is to obtain large enough rate of convergence  
in the fluctuation of attendance Eq.~(\ref{attendance}).  
In these studies we have determined the neighbours by spanning a scale-free 
tree whose nodes denote the agents, and whose links determine the  
neighbours of an agent. With this approach we get fast convergence in  
fluctuations of attendance, because the typical node-to-node distance 
within a scale-free network is short, and thus the information on the 
performance of "good" agents spreads rapidly. Other possibilities  
include, for example, taking the worst fraction of agents and making  
them cross, or letting an agent observe its own performance only,  
and if it continuously decreases, allowing the agent cross its  
strategies. 
 
\section{Results} 
 
\begin{figure} 
\epsfig{file=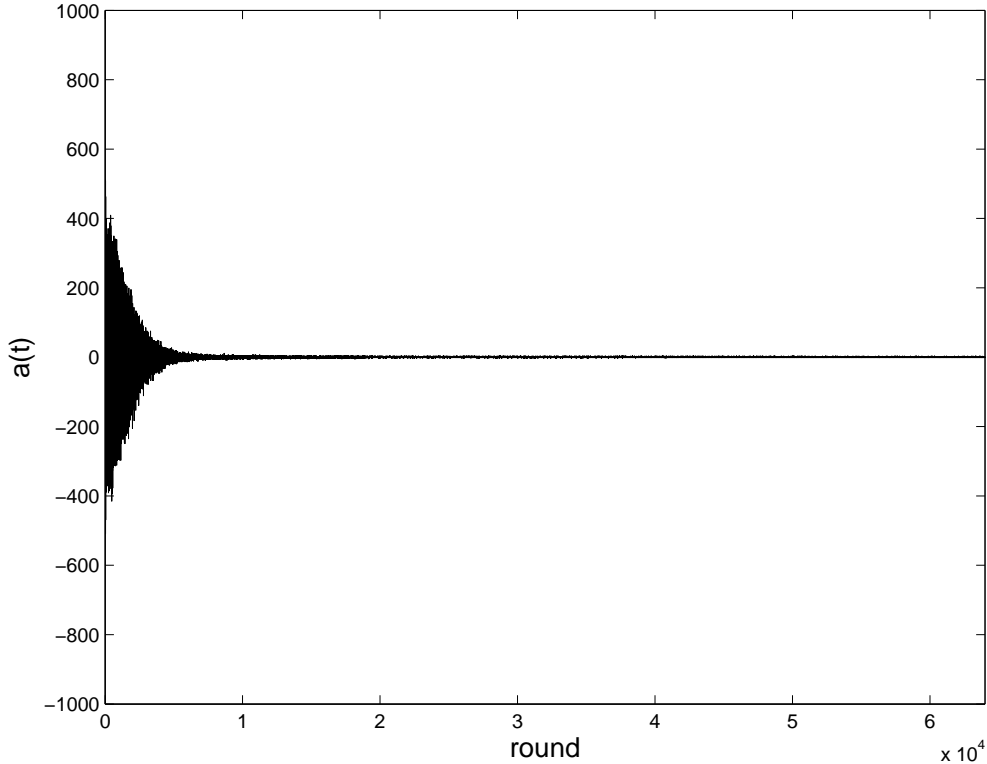,width=5.2in} 
\caption{ The evolution of fluctuations in the EMG during one simulation  
run, with the number of agents $N=1069$, memory length $M=6$,  
number of strategies per agent $S=21$, crossover period $r=2P$, and  
simulation length $C=1000P$, where $P=2^M$. The numbers of minority  
and majority group members (``sellers'' and ``buyers'') eventually  
become as close to each other as possible.  
\label{fig1}} 
\end{figure} 
 
In our numerical simulations we have observed that in our EMG the  
society utility tends to maximize within a wide range of control  
parameter $z=2^M/N$ values, provided that agents are given 
enough strategies at the beginning. In addition we observed that  
agents whose performance is close to each other do not form groups  
in the sense that they would use similar strategies. Also we investigated  
the effects of group decision-making as well as endowing only part  
of the agents with strategy improvement capability. These results  
are explained in detail below. 
 
In what follows, we define the time scale of the simulation in terms of $P$  
rounds, such that $C=const.\times P$. The strategy length $P$ is a  
natural measure of time, since on average, it takes $P$ rounds to  
go through all the components of a particular strategy,  
and thus an agent  
can for each history get response to the success or failure of its choice.  
In Figure \ref{fig1} we show the development of attendance 
Eq.~(\ref{attendance}) during one simulation run. We see that the 
fluctuations start at a high level, but are then rapidly damped 
towards the minimum, indicating that the society utility  
Eq.~(\ref{society_utility}) maximizes. In terms of the trading analogy   
this means that the numbers of sellers and buyers become as close  
to each other as possible, and thus the price of the commodity settles 
down to its equilibrium value. In this simulation run we have used 
$M=6$, $S=21$, $r=2P$ and $C=1000P$ ($64 000$ rounds). The behaviour of  
our EMG differs considerably from that of the BMG, in which the fluctuation  
level would remain high because the control parameter value ($z \approx 0.06$)  
lies within the low-$z$-high-fluctuation region, and because the number  
of strategies $S$ is high \cite{challet4}.  
 
In our EMG model the evolutionary strategy changes mean that agents 
can develop and strive to optimize their strategies with the proven 
crossover method \cite{marko1}, whereas in the BMG model the agents 
are restricted to their original strategies. According to Smith, 
individuals who are striving towards maximizing personal gain 
eventually promote the whole society most effectively.  
This is exactly what happens in our EMG, as the agents do not have any  
explicit rules to lead the society utility to the maximum. Note, however,  
that unlike in the real world, all agents are here equal in their "skills".  
The effect of differing agent abilities will be discussed below. 
 
\begin{figure} 
\epsfig{file=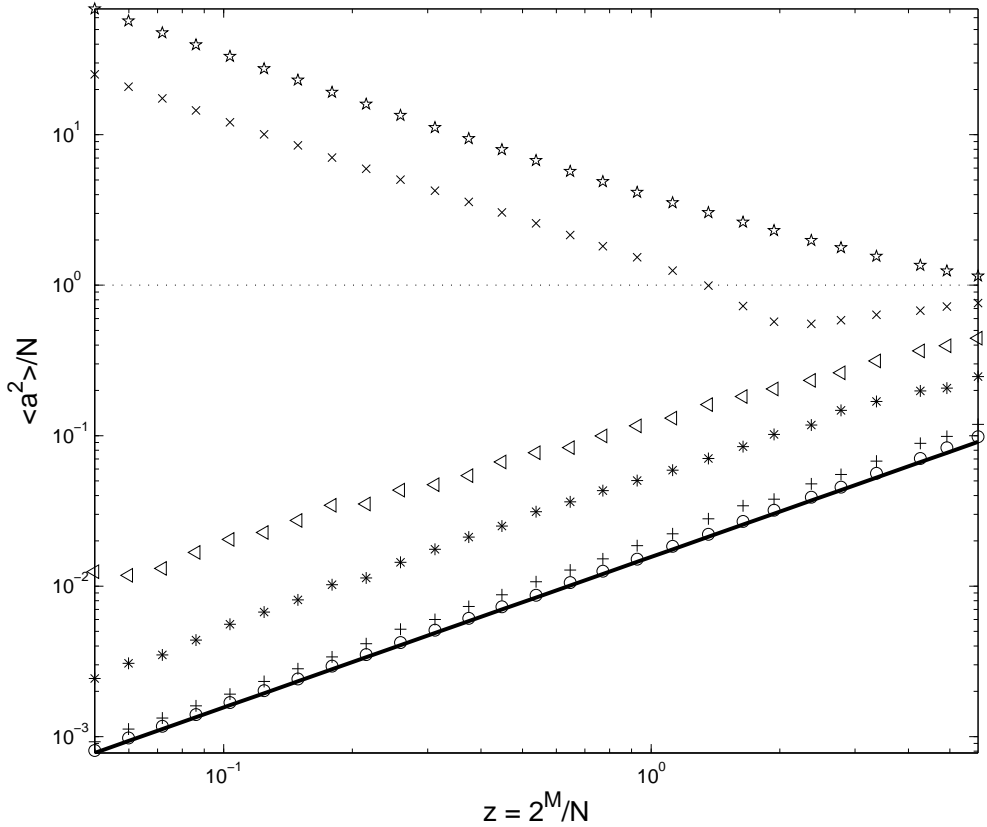,width=5.2in} 
\caption{ Normalized fluctuations versus the control parameter $z = 
  2^M/N$ for $S = 5$ (triangles), $S=8$ (asterisks), $S=13$ (plus-signs), 
  and $S=21$ (circles) in our EMG. For comparison, we plotted also the 
  normalized fluctuations in the BMG for $S=5$ (crosses) and $S=21$  
  (stars). We used $M=6$, $r = 2P$, and $C = 1000P$, and averaged over 
  $200$ estimates. Increasing $S$ leads the normalized fluctuations into the 
  minimum line. 
\label{fig2}} 
\end{figure} 
 
In Figure \ref{fig1} we gave an example of the evolution of attendance 
Eq.~(\ref{attendance}) and maximization of the society utility 
Eq.~(\ref{society_utility}) (minimization of $|a(t)|$) in one particular 
realization of the game with fixed parameter set. Figure \ref{fig2}  
shows the normalized fluctuations of attendance Eq.~(\ref{fluctuation})  
versus the control parameter $z = 2^M/N$, illustrating  how the optimum is  
reached, if enough strategies are given to the agents at the beginning 
of the game. For each point on the curve we have used $M=6$, $r=2P$  
and $C=1000P$, and averaged over $200$ estimates. An estimate for the  
normalized fluctuations Eq.~(\ref{fluctuation}) is calculated using the  
last $1000$ simulation rounds. The number of rounds $r$ after which  
the agents check their neighbours and decide about crossing their  
strategies is two whole periods. In this time an agent gets, on the  
average, two responses for its actions for every history, and thus  
has time to learn which of its strategies perform better than others. 
 
There are two reference lines in Figure \ref{fig2}: the horizontal  
dotted line $\left< a^2 \right> /N = 1$ and the solid line  
$\left< a^2  \right>/N=1/N$. The former indicates the level  
of normalized fluctuations, if the random decision (coin flipping)  
strategy is used, and the latter the minimum value of the normalized  
fluctuations (maximum society utility). The four series below  
the random decision strategy line display normalized fluctuations  
for $S=5$ (triangles), $S=8$ (asterisks), $S=13$ (plus-signs) 
and $S=21$ (circles). All series fall around lines whose  
slopes $\approx 1$ which is the same as that of the minimum  
normalized fluctuations line. As the number of strategies increases,  
the lines start to converge towards the minimum normalized fluctuation  
line. This indicates that the society utility is maximized.  
With $S=21$ the level of normalized fluctuations Eq.~(\ref{fluctuation})  
is very close to its minimum value for all values of the control parameter  
$z$. Thus, in our EMG the society utility Eq.~(\ref{society_utility})  
increases, if we increase the number of strategies per agent $S$. The  
reason for this is that larger initial strategy sets allow more crossover  
combinations, among which the agents can find good ones with higher  
probability. Thus, the strategy set size can be seen as representing  
the initial capabilities of agents, and also sets a limit for  
improvement. If $S$ is too small, it is possible that combinations do  
not include those strategies which finally lead to the society utility maximum. 
 
A remarkable property in the case of $S \geq 21$ is that the 
normalized fluctuation values Eq.~(\ref{fluctuation}) are minimized for 
all simulated $z$. The result is robust and shows how efficient the  
utilized evolutionary method is. For comparison,  
we have also plotted normalized fluctuations for the  BMG for $S=5$ (crosses)  
and $S=21$ (stars). Contrary to our EMG, the normalized fluctuations 
increase, if more strategies are added to agents' initial strategy 
sets \cite{challet4}. Here we can also separate the behaviour of  
normalized fluctuations in the low, middle and high value regions of the  
control parameter $z$. In our EMG, the $S=5$ case is the most inefficient  
compared to the games with higher $S$ in the sense of society utility,  
but still considerably more efficient than the BMG for $S=5$.  In fact the 
difference is huge -- of the order of $\sim100-1000$ -- in the low $z$ region.  
This difference is even bigger for higher values of $S$.  
 
\begin{figure} 
\epsfig{file=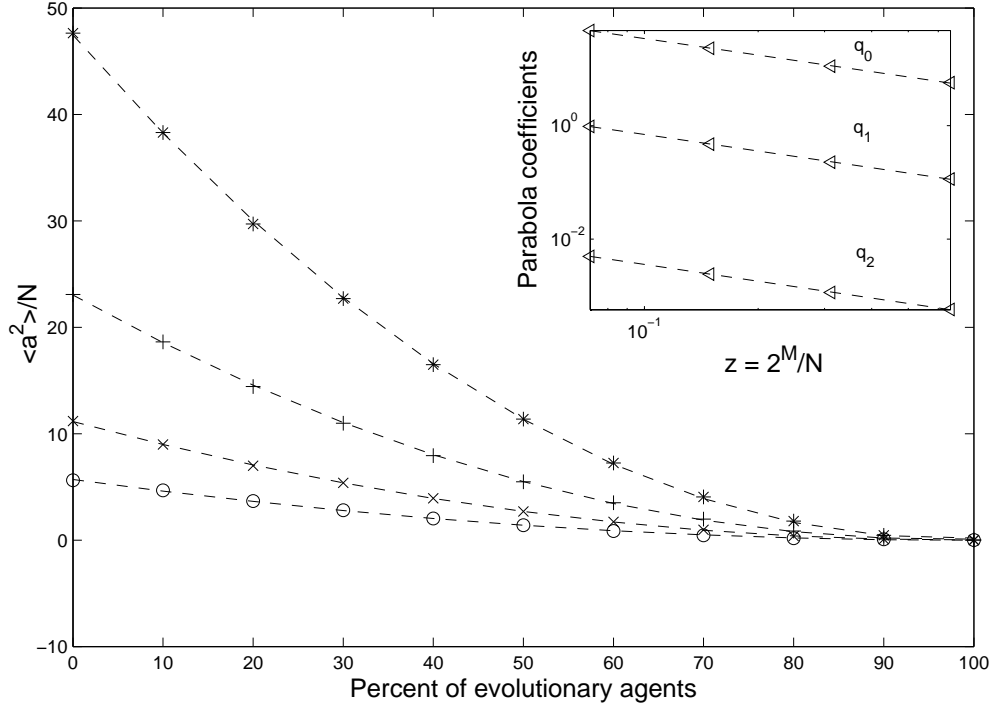,width=5.2in} 
\caption{ Normalized fluctuations versus the percentage of evolutionary 
  agents for $z=0.65$ ($N=99$, $M=6$) (circles), $z=0.31$ ($N=205$, 
  $M=6$) (crosses), 
  $z=0.15$ ($N=429$, $M=6$) (plus-signs), and $z=0.07$ ($N=891$, 
  $M=6$) (asterisks). We used 
  $S=21$, $r=2P$, $C=1000P$, and averaged over $50$ samples. The 
  dashed lines are parabolas fitted to the observations using the 
  minimum square error criterion. The inset shows parabolic coefficients 
  $q_i$ as function of the control parameter, displaying power-law decrease 
  $q_i \propto z^{-1}$. 
\label{fig3}} 
\end{figure} 
 
The assumption that agents are potentially equal in their skills is 
important for reaching the minimum of normalized fluctuations. If a 
fraction of agents is not able to adapt by crossing their strategies, 
the system utility will not reach its maximum value. This can be seen in 
Figure \ref{fig3} where we studied the development of normalized 
fluctuations Eq.~(\ref{fluctuation}) as the percentage of evolutionary 
agents increases for $N=99$, $N=205$, $N=429$, and $N=891$.  
We used $M=6$, $S=21$, $r=2P$, $C=1000P$ for one 
sample run, and averaged over $50$ samples. If none of the agents is 
evolutionary, the level of normalized fluctuations is that of 
the BMG using the same parameters. If all agents are evolutionary,  
the normalized fluctuations Eq.~(\ref{fluctuation}) are minimized as in 
Figure \ref{fig2}. Between these two extremes the normalized 
fluctuations decrease monotonically as the percentage of evolutionary 
agents increases.  
We found that the results are best described by parabolic  
decrease ($<a²>/N = q_2x^2-q_1x+q_0$) of fluctuations as function  
of the fraction $x$ of evolutionary agents [see dashed lines in  
Figure \ref{fig3}]. The values of the coefficients of parabolas seem  
to obey a power-law $q_i \propto z^{-1}$, as indicated by the inset  
in Figure \ref{fig3}. 
 
\begin{figure} 
\epsfig{file=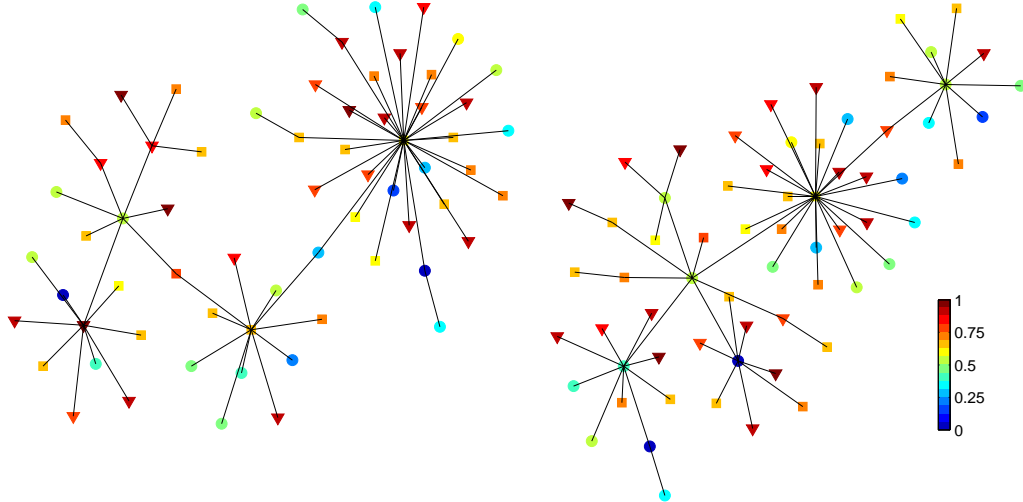,width=2.7in, angle = 270} 
\caption{ The minimum spanning tree (left) and maximum spanning tree (right)
of pairwise Hamming distances 
  between strategies used at the end of simulation. $N=65$, $M=6$, 
  $S=21$, $r=2P$, and 
 $C=3000P$. Performances are observed from the last $100P$ rounds.  
\label{fig4}} 
\end{figure} 
 
Returning to the setting where all agents are evolutionary,  
we studied whether agents whose performance is 
close to each other form groups within which agents use similar 
strategies. If such groups exist, there might be particular strategies 
in the whole strategy space which are preferred compared to others, 
and agents who use the same or similar strategies would perform about 
equally successfully. Here we measure the similarity of two strategies 
$k_1$, and $k_2$ with the Hamming distance 
 
\begin{equation} 
d_{k_1k_2} = \frac{\sum_{i=1}^P |k_1(i)-k_2(i)|}{P}. 
\label{hd} 
\end{equation} 
 
In order to study the formation of groups we have simulated our EMG  
with $N=65$, $M=6$, $S=21$, $r=2P$, $C=3000P$, and observed the performances  
of the agents from the last $100P$ rounds of the simulation. The large  
number of simulation rounds guarantees that agents make a sufficient  
number of crossovers and that the evolution of their strategy pools  
has more or less stopped. In our simulations such a stabilization happens 
often in $C=1000P$ rounds. At the end of the simulation run we take  
notice of the used strategy of each agent and calculate the Hamming distance  
for all possible strategy pairs between agents. 
 
In order to visualize the clustering of either winning or losing strategies  
we have used the minimum/maximum spanning tree methods formed by using pairwise  
Hamming distances. The minimum/maximum spanning tree is the shortest/longest
 tree graph which can be spanned between the nodes \cite{algorithm}.
If some strategy pairs resemble each other, their Hamming distances are small, and thus these  
distance pairs of the whole Hamming distance matrix with $N(N-1)/2$ elements  
will be extracted for the minimum spanning tree, whereas in the maximum spanning
tree, interconnected strategies are far from each other in the strategy space.
In Figure \ref{fig4} we show the resulting  
spanning trees -- minimum on the left, maximum on the right --
which are coloured according to the performance of an agent, scaled into 
the range of $[0,1]$ such that red stands for the best performing agents while 
blue for the worst performing agents. A more coarse-grained division of the 
performances of agents is indicated by three symbols: triangles for the best  
performing third of the agents, circles for the worst performing third  
and squares for the agents whose performance is in the middle of these two.  
  On one hand, the minimum spanning tree
shows that there is no clear clustering of strategies for similarly performing agents,  
because if such clusters existed, these would be seen as similarly coloured  
clusters of agents. On the other hand, the best-performing agents are 
typically connected to less well performing agents in the maximum
spanning tree. This indicates that the strategies of well performing agents
tend to be far from the strategies of less well performing
agents. Furthermore, because well performing agents are 
never connected to other well performing agents in the maximum
spanning tree, their strategies can
not be very far from each other in the strategy space. 
In addition it is worth mentioning that distribution of the performances of 
agents turned out to be approximatively Gaussian.
 
\begin{figure} 
\epsfig{file=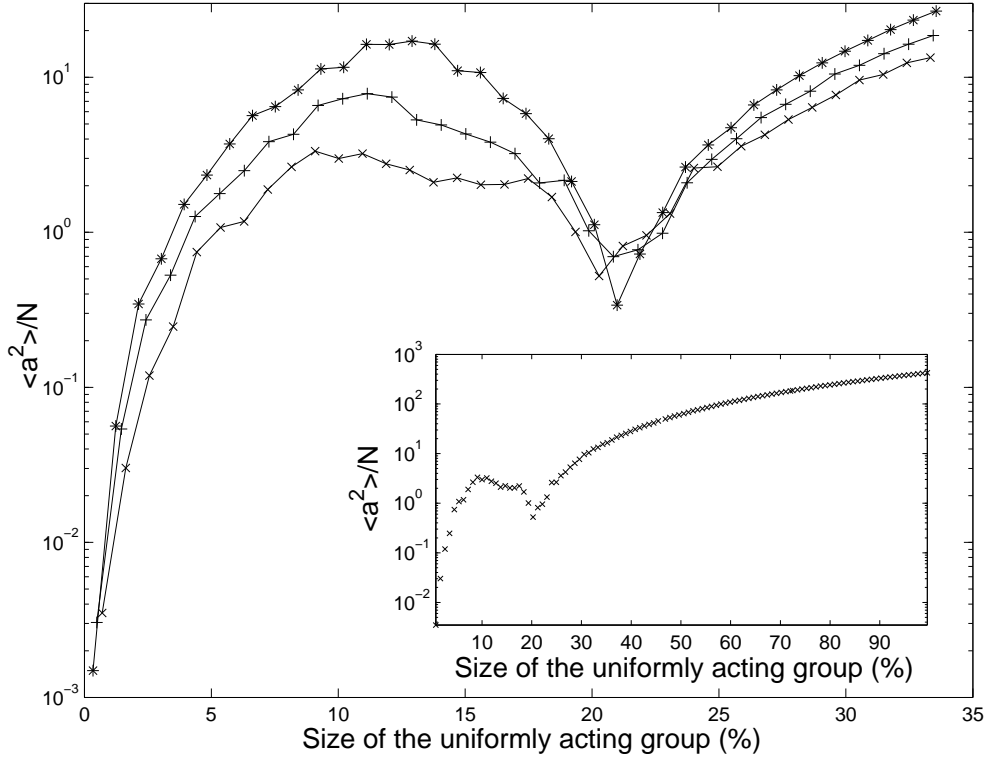,width=5.2in} 
\caption{ Normalized fluctuations versus the group size of uniformly 
  acting agents for $N=891$ (asterisks), $N=619$ (plus-signs), and $N=429$ 
  (crosses). We used $M=6$, $S=21$, $r=2P$, and $C=1000P$, 
  and averaged over $50$ sample runs. For group sizes over $\sim 21$ percent 
  of the whole population the fluctuations grow monotonically, as the 
  group size is enlarged, and  finally reach $N$ for $100$ 
  percent. As an example, the full range for $N=429$ is shown in the 
  inset. However, group sizes below $\sim 21$ show counter-intuitive 
  behaviour as there are local maxima and minima on the curves.}     
\label{fig5} 
\end{figure} 
 
So far, we have considered the agents in the game as individuals 
making independent decisions. To investigate effects of the presence 
of a uniformly acting group, ``cartel'', on the outcome of the game, 
we have developed a variation of our EMG where  
a certain fraction of agents make a group decision and its 
members always obey this decision in their actions. 
 The group decision is done in such a way that 
every round the agents of the ``cartel'' make tentative minority group 
votes according to their strategies, and then the ``cartel''  
decides the final minority group choice for all its members based  
on these votes. This is done taking into account the 
``minority wins''--aspect of the game, such that the final group 
decision is the one for which the \textit{minority} of the agents voted. 
All agents in the group then act according to this decision. 
 
In Figure \ref{fig5} we show the normalized fluctuations as the size  
of the uniformly acting group increases for $N=891$ (asterisks), 
$N=619$ (plus-signs), and $N=429$ (crosses). We used $M=6$, $S=21$,  
$r=2P$, $C=1000P$, and averaged over $50$ samples. For very  
small group sizes, i.e. $0-2$ percent of the whole population, the system  
does not suffer from a big loss in the society utility, but as the group  
size grows, the normalized fluctuations increase until they reach a  
local maximum at group size of about  $\sim 10-13$ percent 
of the agent population. After this, there 
is as yet unexplained decrease in the fluctuation values. As the 
group size further increases over $\approx 21$ percent, the fluctuations  
begin increasing monotonically with the group size, and finally reach 
$N$ in accordance with the Eq.~(\ref{fluctuation}). The inset in 
Figure \ref{fig5} shows an example of this growth in the case 
$N=429$. The local peaks and the minima of the curves are rather  
counter-intuitive, as group decisions mean that part of the agents 
are forced to vote similarly. Therefore, one could expect that 
the normalized fluctuations always increase with the group size. 
One possible explanation for the minimum might be that as the group size 
increases, while still remaining below some certain limit, the group  
is better able to predict the minority choice using statistics 
provided by its own members. 
 
\section{Conclusions} 

To summarize, we have presented an evolutionary modification to the original 
Minority Game model \cite{challet1}, where individual agents are  
capable of learning from the outcomes of their past decisions and  
changing their strategies accordingly. This modified game leads  
to a stable situation where the majority and minority group sizes  
become almost equal for a wide range of simulation parameters.  
Typically, this happens for all possible histories, and each agent's  
choice for each history corresponds to one of the game's pure-strategy Nash 
equilibria. This phenomenon can be seen as an example of  
self-organization in a complex evolutionary system \cite{arthur,lendaris},  
where the evolution is driven by competition among agents. The optimized  
state emerges as a result of the selfish pursuits of individual  
agents. If the game is viewed as a toy model of market economy, the 
equal group sizes mean that the amounts of buyers and sellers of commodity 
are identical, which drives the commodity price to its equilibrium value, 
and society utility to its maximum. This is analogous to the  
``Invisible Hand'' -effect predicted by Adam Smith, stating that 
selfishly-acting individuals who are not actively concerned with 
the welfare of the whole society still eventually reward the whole 
society in an optimal way.  
 
In principle, a similar equilibrium state of minimal fluctuations  
could be reached by dividing the agents into two almost equal-sized  
groups, A and B. Then, group A's minority choice would always be 1  
(or -1), and group B's choice the opposite. Our studies show that  
this does not happen in the EMG and thus clear clusters of agents 
utilizing the same or similar strategies do not form. This is intuitively
quite evident, as using strategies different from those of other 
agents increases the probability of being in the minority group. 
Hence, the agents' strategies tend to move away from each other 
in the strategy space, rather than converge. In addition,
the strategies of well-performing agents tend to be far from
the strategies of agents with worse performances. 
 
We have also observed that the optimal outcome of the society is 
reached only under idealized conditions, where all the agents are 
equally capable of modifying their actions -- a condition which 
is rarely met in real-world economic systems. If only part of the 
agents are allowed to evolve their strategies, the fluctuations  
do not reach a minimum. Furthermore, we have found that the 
fluctuations decrease quadratically as a function of the fraction 
of evolving agents. As for the homogeneity of the agent population is
concerned, we have also investigated the effect of simulated uniformly 
acting  ``cartels''.  The expected result emerges such that if there is 
a cartel in the society its utility is not maximized, i.e. true price 
equilibrium is not reached. Surprisingly we also find that introducing 
a uniformly acting cartel does not lead to steadily increasing amplitude 
of the fluctuations as would be expected if there is a cartel in a 
game where the agents chose their side randomly. Instead of 
a steady increase in fluctuations we observe a local minimum 
when the uniformly acting cartel includes $\sim 20$ percent of the agents, 
for several system sizes studied. One possible reason might be that 
the minima arise from a combination of two factors. On one hand, the 
agents involved in the uniformly acting cartel are able to better 
estimate the winning side when the cartel size becomes large enough.
On the other hand, as the cartel size grows, the fraction of agents, 
which are free to choose their actions and thus counter the effect of 
the cartel decreases, leading to increased fluctuations. 

Previously, the existence of minimal fluctuations in minority games  
has been discussed and observed by several authors for systems of agents using
probabilistic rules for strategy selection \cite{bot,marsili,reents,cav}. 
In contrast, in our model, the mechanisms for strategy selection are
deterministic and do not include probabilities. In the case of
pure strategies, in Ref.~\cite{marsili} the authors show that minimum 
fluctuations can be achieved if the agents are allowed to remove their 
contribution from the outcome of the game. In other words the agents are 
allowed to subtract their choice from the attendance, Eq.~(\ref{attendance}), 
used in determining the change in their personal utility function, 
$\Delta u_i = -\text{sgn}(a-\eta\sigma_i)$. However, in our model the 
strategy score updating rules are the same as those in the original 
BMG \cite{challet1}, without any extensions or alterations.  
Nevertheless, the state of minimum fluctuations is typically reached 
in our game, due to the effectiveness of the genetic crossover 
mechanism.

\section{Acknowledgments} 
This research was partially supported by the Academy of 
Finland, Research Centre for Computational Science and Engineering, 
project no. 44897 (Finnish Centre of Excellence Programme 2000-2005). 
We want to thank professor J\'{a}nos Kert\'{e}sz from Budapest University 
of Technology and Economics for his ideas related to the use of 
Minimum Spanning Tree method in this research.

\end{document}